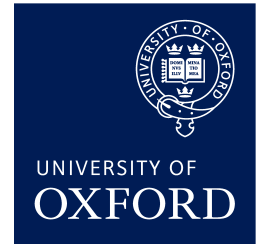

Konrad Kollnig and Reuben Binns
Department of Computer Science, University of Oxford

13 May 2022

**Response: 'GDPR and the Lost Generation of Innovative Apps'**

A recently published pre-print titled 'GDPR and the Lost Generation of Innovative Apps'[1] observes that a third of apps on the Google Play Store disappeared from this app store around the introduction of the GDPR in May 2018. The authors deduce *'that GDPR is the cause'*. The effects of the GDPR on the app economy are an important field to study. Unfortunately, the paper currently lacks a control condition and a key variable. As a result, the effects on app exits reported in the paper are likely overestimated, as we will discuss. We believe there are other factors which may better explain these changes in the Play Store aside from the GDPR.

This response is structured as follows. We first review key statistics around apps on the Google Play and Apple App Store, and highlight contradictions with the claims by the authors of the paper. Specifically, we find that the **Apple App Store has not seen a similar drop in apps as on Google Play** and **that the EU makes up less than 20% of revenues**. We then highlight that **Google first began removing outdated apps from its Play Store and improving its review process from 2017**, a key variable that was not considered by the paper. It has been widely understood in the privacy community that Google started conducting its first and last major purge of apps on its app store from December 2017, and that this led to the decline in the number of apps. This point has unfortunately been missed by the authors. Rather than being a study on *'app exits'* and the impact of the GDPR (as the authors claim), the study rather analysed Google's *removals of apps* and the impact of Google's content moderation on apps at the time.

In sum, without drawing on an extensive set of additional data, it will be difficult to keep up key claims made in the paper. In particular, it will be near to impossible to disentangle the impact of the GDPR from the impact of Google's actions on low-quality apps on the Play

---

[1] Janßen et al. (2022). GDPR and the Lost Generation of Innovative Apps. *NBER Working Paper Series*. https://www.nber.org/papers/w30028

Store without further data provided by Google. This extends to both observations about the drop in app count as well as the claimed negative impact on innovation.

Given the difficulty in analysing key questions in the app platform economy – like the impacts of the GDPR, we conclude that the gatekeepers need to allow more transparency for the interested public, in particular through disclosures about privacy-related activities[2].

## App Statistics

We first discuss relevant key statistics around the Google Play and Apple App Store.

**The number of apps on Google Play substantially dropped in 2018**

Let us first consider the number of apps on the Google Play Store over time. It was indeed the case that the number of apps saw a heavy decline around the introduction of the GDPR (i.e. 25 May 2018). This decline actually already started on 29 March 2019, when the number of apps suddenly dropped by 116,654 apps – within a single day and more than a month before the GDPR came into force. Until the end of April 2018, the total number of apps further declined by 0.5 million and then suddenly reached an equilibrium for a month. From 19 May (i.e. 6 days before the GDPR came into force), the app count started to fall again rather by a further 0.75 million until October. This is also visualised in Figure 2, which suggests that there were 4 different episodes of app decline in 2018.

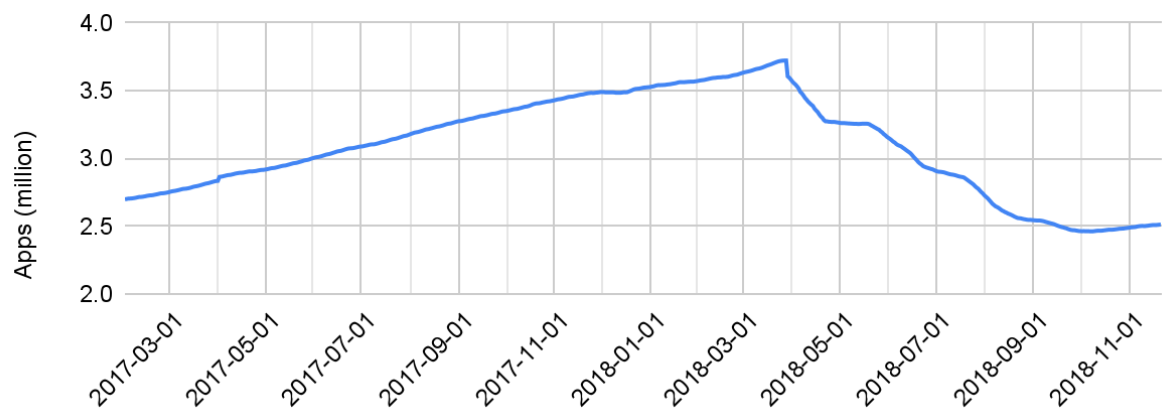

Figure 1. Apps on the Google Play Store over time. Source: AppBrain [3]

---

[2] Van Hoboken and Ó Fathaigh (2021). Smartphone platforms as privacy regulators. *Computer Law & Security Review, 41*. https://doi.org/10.1016/j.clsr.2021.105557

[3] AppBrain (2019). Android and Google Play statistics. https://web.archive.org/web/20190301002622/https://www.appbrain.com/stats

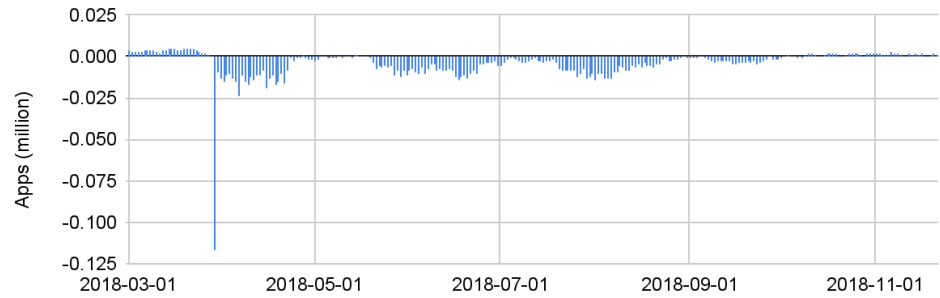

Figure 2. Change in Total Apps Per Day on Google Play. Source: AppBrain [4]

The authors claim that the *'sharp spike in exit [i.e. drop in number of apps on Google Play] as GDPR takes effect is prima facie evidence that GDPR is the cause.'* Close inspection, however, shows that the pattern with which the count of apps declined is at least somewhat inconsistent. Our observations raise some initial doubt over the claim by the authors that it was clearly the GDPR causing this drop. Additionally, the number of app downloads actually increased by about 16% from Q4 2017 to Q4 2018[5], which seems to contradict the claim of the authors that there was a major hit to the app economy at the time.

**Most revenues are generated outside of jurisdiction of GDPR**

If the GDPR has affected apps' viability *'substantially'* (and *'induced the exit of about a third of available apps'*), then one would expect that the EU is among the most important drivers for revenues and for app downloads. Neither is true.

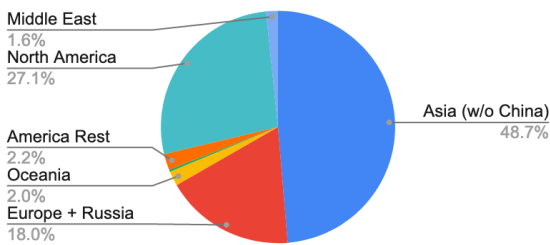 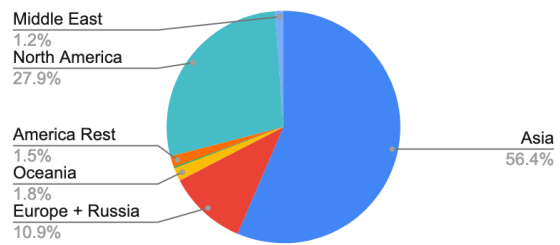

Figure 3. App revenues by country in 2018. Source: SensorTower [6]

---

[4] See footnote 3.
[5] Sensortower (2019). Q4 and Full Year 2018: Store Intelligence Data Digest.
https://sensortower-itunes.s3.amazonaws.com/Quarterly+Reports/Sensor-Tower-Q4-2018-Data-Digest.pdf
[6] SensorTower (2019). 5-Year Market Forecast: App Store and Google Play Spending Will Grow 120% to Reach $156 Billion by 2023.
https://sensortower.com/blog/sensor-tower-app-market-forecast-2023

If we look at revenues by region, we see that Europe (and hence also the EU) makes up a relatively small share of overall revenues. On Google Play, less than 18% of revenue is generated in Europe, compared to less than 11% on the App Store. Most revenues are generated in regions to which the GDPR does not apply, i.e. North America and Asia. In light of this, it is surprising that the authors of the above-mentioned paper find a substantial effect of the GDPR on Google Play. While one could argue that there are spillover effects between jurisdictions, the relatively small share of EU countries among revenues casts some doubt over this argument.

Looking at the countries with the most downloads, the top 10 countries on the Google Play Store turn out to be all non-EU countries[7]. Again, the GDPR does not apply to any of these countries.

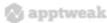

| | Country | Total downloads | | Country | Total downloads |
|---|---|---|---|---|---|
| 1st | India | 5,066,495,300 | 1st | United States | 2,098,141,490 |
| 2nd | Brazil | 1,726,370,000 | 2nd | Japan | 346,188,460 |
| 3rd | Indonesia | 1,483,543,690 | 3rd | United Kingdom | 251,664,260 |
| 4th | United States | 1,235,055,930 | 4th | Russia | 247,054,143 |
| 5th | Russia | 868,051,720 | 5th | Brazil | 164,016,360 |
| 6th | Mexico | 790,091,205 | 6th | France | 161,071,170 |
| 7th | Vietnam | 594,005,610 | 7th | Vietnam | 154,635,810 |
| 8th | Turkey | 591,268,590 | 8th | Germany | 142,297,920 |
| 9th | Egypt | 419,464,880 | 9th | South Korea | 419,464,880 |
| 10th | South Korea | 362,930,140 | 10th | Turkey | 137,371,630 |

Figure 4. App downloads by country in 2019. Source: apptweak[8]

**The number of apps on the App Store stayed almost constant in 2018**

If the GDPR had indeed had a strong effect on the viability of apps, it should show up, too, in the Apple App Store. It turns out that there was no such similar effect on the Apple App Store. Between Q1 and Q2 of 2018, the app count remained constant. Between Q2 and Q3, the number of apps decreased from 2.03 to 1.96 million, following a downward trend since

---

[7] Apptweak (2019). Countries with Most App Downloads.
https://www.apptweak.com/en/aso-blog/infographic-countries-with-most-app-downloads
[8] Apptweak (2019). Countries with Most App Downloads.
https://www.apptweak.com/en/aso-blog/infographic-countries-with-most-app-downloads

2016. The app count on the App Store had already fallen by 8.14% before the GDPR came into force, and by another 9.85% afterwards over the following quarters. In other words, the App Store lost about a tenth of apps before and after the GDPR, rather than a third of apps as on Google Play. There lacks evidence to explain these observations and to attribute them to the GDPR. The number of apps on the app stores does thus not seem to serve as a robust measure to assess the impact of the GDPR. Despite this, the *'quarterly data on each app's availability'* forms a key facet of the above-mentioned paper, as also explained in the previous paragraph. Because the number of apps has changed, these authors deduce that the GDPR was the root cause. This observation is not confirmed in the natural control group, i.e. the Apple App Store.

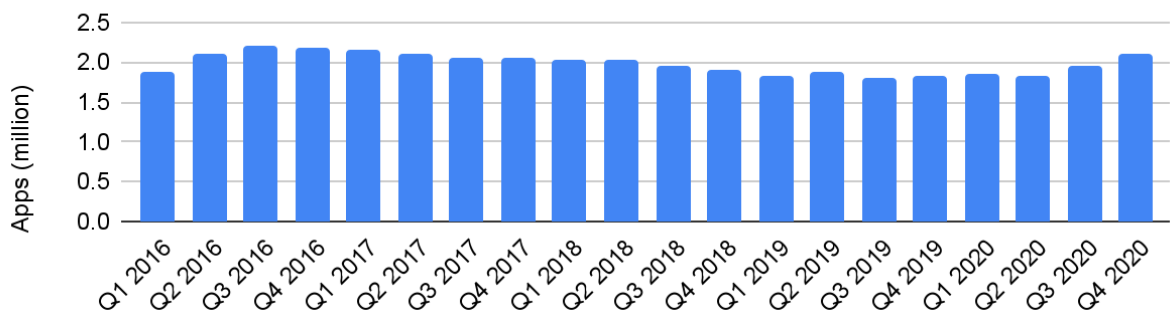

Figure 5. Apps on the Apple App Store over time. Source: Statista [9]

# Causes of app decline

The drop in apps on the Google Play Store has long been noticed by those who watch the industry. From the above, it seems unlikely that the reason was the GDPR because the Apple App Store showed no similar drop but is much more important for EU users.

We now discuss likely causes for a decline of apps on Google Play, but not on the Apple App Store.

**Explanation 1: Removal and hiding of poor quality apps on Google Play**

A commonly mentioned explanation for the drop in apps on the Google Play Store is not the GDPR, but rather a change in Google's machine learning algorithms. Contrary to Apple, Google does traditionally not review every app manually and instead relies on machine learning algorithms. As we see from lively discussions around the impact of such algorithms on society, these algorithms have generally been improving significantly in terms of quality over recent years. The algorithms that ensure the review of apps on the Google Play Store are no exception[10]. Indeed, as we've experienced from our own research of app privacy at scale[11], a significant number of apps are almost exact clones of popular apps. Often they just steal the code from a popular app which you have to pay for, slightly rebrand it, and release it with added adtech trackers in it; this destroys the business model of many small app developers pursuing a subscription model[12]. So, many of these missing apps might just be clones and other low-quality apps that are now being picked up by Google's improved review processes from 2017 onwards.

---

[9] Statista (2022). Number of available apps in the Apple App Store from 1st quarter 2015 to 1st quarter 2022.
https://www.statista.com/statistics/779768/number-of-available-apps-in-the-apple-app-store-quarter/

Beyond the changes to the app review process, Google has also been changing its search algorithm on the Play Store so as to suggest better apps to users. As a result, some apps suddenly saw their download counts drop; this has been especially reported in 2018[13].

Both aspects, the changed review and search methods, might be important factors to explain that 'in the quarters following implementation [of the GDPR], entry of new apps fell by half'.

---

[10] TechCrunch (2018). Google says it removed 700K apps from the Play Store in 2017, up 70% from 2016. https://techcrunch.com/2018/01/30/google-says-it-removed-700k-apps-from-the-play-store-in-2017-up-70-from-2016/ and Digital Information World (2022). Google Play Store dropped 1 million apps, but it's nothing to worry about. https://www.digitalinformationworld.com/2022/04/google-play-store-dropped-1-million.html

[11] Kollnig et al. (2021). Before and after GDPR: tracking in mobile apps. *Internet Policy Review, 10*(4). https://doi.org/10.14763/2021.4.1611 and Binns et al. (2018). Third Party Tracking in the Mobile Ecosystem. *ACM WebSci 2018*. https://dl.acm.org/doi/10.1145/3201064.3201089 and Kollnig et al. (2022). Are iPhones Really Better for Privacy? A Comparative Study of iOS and Android Apps. *Proceedings on Privacy Enhancing Technologies, 2022*(2). https://doi.org/10.2478/popets-2022-0033

[12] Ekambaranathan et al. (2021). "Money makes the world go around": Identifying Barriers to Better Privacy in Children's Apps From Developers' Perspectives. *ACM CHI 2021*. https://doi.org/10.1145/3411764.3445599 and Android Authority (2021). Google's new rules appear to be successfully purging the Play Store. https://www.androidauthority.com/google-play-store-purge-2732281/

[13] Google (2018).
Improving discovery of quality apps and games on the Play Store.
https://android-developers.googleblog.com/2018/06/improving-discovery-of-quality-apps-and.html
and Variety (2018). Google Changes 'Discovery Algorithm' on Google Play, Leading to Panic From Some Devs.
https://variety.com/2018/gaming/news/developers-sudden-drop-downloads-on-google-play-store-1202861850/ and Unity Forums (2018). Sudden drop in number of daily installs on Google Play Store.
https://forum.unity.com/threads/sudden-drop-in-number-of-daily-installs-on-google-play-store.537467/
and Quora (2018). Why did the number of my app downloads decreased rapidly on Google Play on June 22?
https://www.quora.com/Why-did-the-number-of-my-app-downloads-decreased-rapidly-on-Google-Play-on-June-22

**Explanation 2: *Removal* of outdated apps on Google Play from 2018**

A strong impact on the download count of apps will have been due to the fact that Google first began removing outdated apps from 2018[14]. In December 2017, Google announced that all apps would soon have to meet a minimum API level (i.e. the version of apps targeted during development). Apps that had not been updated by the deadline were removed from the Google Play Store. Naturally, this had the effect that the number of apps substantially declined afterwards. The Google Play Store had been in operation for 10 years then and had never been cleaned up from outdated apps at such scale before.

Additionally, it had already been documented since 2017 that Google was starting a major purge of the Google Play Store of apps that were not compliant with its data protection and privacy rules[15]. For instance, apps that accessed certain sensitive personal information (e.g. the Android Advertising Identifier) but failed to provide a privacy policy were removed over time. While this purge of apps was first started in February 2017, the app statistics on apps on the Google Play Store (see Figure 1) suggest that this purge was only conducted in 2018 (because there was no noticeable change in 2017).

This is also documented in the Wikipedia entry of the Google Play Store: 'By 2017, Google Play featured more than 3.5 million Android applications. After Google purged a lot of apps from the Google Play Store, the number of apps has risen back to over 3 million Android applications.'[16]

This evidence suggests that it often was not app developers making the decision to quit the Play Store, but rather Google removing these apps.

By contrast, Apple has constantly been removing outdated apps from its App Store – since its inception. The way that they do that is by requiring every app developer to renew their membership once per year. As part of this, developers must pay a 99$ membership fee.

---

[14] XDA Developers (2017). Play Store will Require New and Updated Apps to Target Newer API Levels and Distribute Native Code with 64-bit Support. https://www.xda-developers.com/play-store-updated-requirements-api-level-64-bit/ and primetel (2022). Google will remove outdated apps from Google Play store. https://primetel.com.cy/google-will-remove-outdated-apps-from-google-play-store-3056 and Stackoverflow (2018). Target API level requirement from late 2018. https://stackoverflow.com/questions/50986651/target-api-level-requirement-from-late-2018

[15] The Next Web (2017). Millions of apps could soon be purged from Google Play Store. https://thenextweb.com/news/millions-apps-soon-purged-google-play-store and Stackoverflow (2018). I have added privacy policy to play store and my application. It's been 10 hours my application didn't publish again https://stackoverflow.com/questions/52533830/i-have-added-privacy-policy-to-play-store-and-my-application-it-s-been-10-hours

[16] Wikipedia (2022). Google Play: Android applications. https://en.wikipedia.org/w/index.php?title=Google_Play&oldid=1086799566#Android_applications

Google Play only requires a $25 one-time registration fee. Because of this apps could theoretically stay on the Play Store forever – until Google first started removing outdated apps in 2018. This was 10 years after the launch of Google Play in 2008, and thus the number of apps removed was rather high.

The authors acknowledged that policy changes by Google might have affected Play Store statistics, but found that these *'policy changes occurred either substantially before, or long after, GDPR took effect'* and concluded that they *'cannot explain the exit spike'*. This suggests that they missed an important policy change, namely Google's app cleanup.

Rather than apps deciding to *exit* the Play Store at scale, Google *removed* them at great scale. This is unfortunately not reflected nor modelled in the paper.

## Conclusions

In sum, the above-mentioned paper fails to provide sufficient evidence and to prove causality. The paper lacks a control condition, and disregards the fact that the Apple App Store has not seen a similar drop in the number of apps at the same time.

The reason for the sudden drop in apps was likely less due to the GDPR but rather the fact that **Google first started removing outdated apps in 2018**. In 10 years of operating the Play Store, many outdated and low-quality apps had accumulated on the store and were purged in 2018. Traditionally, many such low-quality and copycat apps used to be on the Play Store, in part because Google used to rely entirely on automated app review, and not manual review as on the App Store. This is increasingly being addressed by Google – a positive improvement for end-users. Improved search and review algorithms on the Google Play Store have further increased the quality of apps, leading to fewer apps joining the Play Store. In other words, **the model of the paper lacks a key variable**, namely the number of removals of apps by Google, and struggles to distinguish the two effects (i.e. Google's activities and app developers' choices). Instead of being a study on *'app exits'* and the impact of the GDPR (as the authors claim), the study rather analysed Google's *removals of apps* and the impact of content moderation on malicious and copycat apps.

Given the difficulty in analysing key questions in the platform economy – like the impacts of the GDPR, we conclude that app platform gatekeepers need to allow more transparency for the interested public, in particular by providing statistics on their policing of the app ecosystem. This has previously been underlined by Van Hoboken and Ó Fathaigh who found that Google and Apple increasingly act as important regulators of data protection and privacy, but with limited regulation, oversight, and accountability[17]. To increase transparency, these authors argued for mandatory disclosures about the privacy-related activities of smartphone platforms – as a minimally invasive but realistic intervention.

Lastly, it is important to acknowledge that we do not argue against the fact that the GDPR has affected digital ecosystems substantially and might have made some apps unviable. Rather, we think that this needs to be supported with good evidence and to understand the wider context, which the said paper fails to accomplish.

---

[17] See footnote 2.